\documentclass{aa}  
\usepackage{graphicx}
\usepackage{txfonts}
\usepackage{natbib}

\begin{document}
   \title{Minimization of non common path aberrations at the Palomar
     telescope using a self-coherent camera}

\author{R. Galicher\inst{1}, P. Baudoz\inst{1}, J.-R. Delorme\inst{2},
  D. Mawet\inst{2}, M. Bottom\inst{3}, J. K. Wallace\inst{3}, E. Serabyn\inst{3}, C. Shelton\inst{3}}
\institute{Lesia, Observatoire de Paris, PSL Research University, CNRS, Sorbonne Universités, Univ. Paris Diderot, UPMC Univ. Paris 06, Sorbonne Paris Cité, 5 place Jules Janssen, 92190 Meudon, France\\
\email{raphael.galicher at obspm.fr}
\and {Department of Astronomy, California Institute of Technology,
  Pasadena, CA 91125, USA}
\and {Jet Propulsion Laboratory, California Institute of Technology,
  Pasadena, CA 91109, USA}}
  \date\today

 
  \abstract
      {The two main advantages of exoplanet imaging are the discovery
        of objects in the outer part of stellar systems --
        constraining the models of planet formation --, and its
        ability to spectrally characterize the planets -- giving
        information on their atmosphere. It is, however,
        challenging, because exoplanets are up to~$10^{10}$ times
        fainter than their star and separated by a fraction of
        arcsecond. Current instruments like SPHERE/VLT or 
        GPI/Gemini detect young and massive planets, because
        they are limited by non-common path aberrations (NCPA) that
        are not corrected by the adaptive optics system. To probe
        fainter exoplanets, a new class of 
        instruments capable of minimizing the NCPA is needed. One
        solution is the self-coherent camera (SCC) focal plane
        wavefront sensor, whose performance was demonstrated in
        laboratory attenuating the starlight by factors up to
        several~$10^8$ in space-like conditions at angular separations
        down to~$2\,\lambda/D$.}
   {In this paper, we demonstrate the SCC on the sky for the first
     time.}
   {We installed an~SCC on the stellar double
     coronagraph (SDC) instrument at the Hale telescope. We
     used an internal source to minimize the NCPA that limited
     the vortex coronagraph performance. We then compared to the
     standard procedure used at Palomar.}
   {On internal source, we demonstrated that the~SCC improves
     the coronagraphic detection limit by a factor
     between~$4$ and~$20$ between~$1.5$ and~$5\,\lambda/D$. Using
     this~SCC calibration, the on-sky contrast is
     improved by a factor of~$5$ between~$2$ 
     and~$4\,\lambda/D$. These results prove the ability of the~SCC to
     be implemented in an existing instrument.} 
   {This paper highlights two interests of the
     self-coherent camera. First, the~SCC can minimize the speckle
     intensity in the field of view especially the ones 
that are very close to the star where many exoplanets are to be
discovered. Then,
     the~SCC has a~$100\%$ efficiency with science time as each image
     can be used for both science and~NCPA minimization.}
   \authorrunning{Galicher et al.}
   \titlerunning{SCC at Palomar}
   \maketitle
%

\section{Introduction}
Imaging of exoplanets is one priority for astronomers because
it is the only technique that can discover long orbital period planets
and that enables the spectral characterization of their
atmospheres. It is very challenging as the planets are
$10^4$
to~$10^{10}$ dimmer than their host star and at a fraction of
arcsecond from their host star. Many coronagraphs were
proposed to reduce the star diffraction pattern without changing the
exoplanet image~\citep{snik18}. Several are installed on the~8m class
telescopes in instruments like~SPHERE~(Beuzit et al., submitted)
and~GPI~\citep{macintosh14} that were built to discover exoplanets by
imaging. The coronagraphic images produced by these instruments enable
the detection of planets that are up to~$\sim10^6$ times fainter than
their star. Such performance is far from the best performance reached
in laboratory with attenuation of the starlight by a factor
of~$10^{9}$ to~$10^{10}$~\citep{baudoz18b,lawsonPR13}. That is because wavefront
aberrations upstream the coronagraph can be measured and minimized
down to a few picometers~rms in laboratory using the technique of focal
plane wavefront sensing and control like the pair-wise
technique~\citep{giveon11} coupled with electric field
conjugation~\citep{giveon07c}, speckle nulling~\citep{Borde06} or the
self-coherent camera~\citep{galicher08}. Behind a
ground-based telescope, it is more complicated to control the aberrations
because they are not static but quasi-static with respect to the
exposure times used to record the coronagraphic image. As a
consequence, even using focal plane wavefront control like speckle
nulling~\citep{martinache14,bottom16b} or
EFC~\citep{cady13,matthews17}, the level of aberrations is
about~$10$\,nm rms limiting the starlight attenuation to factor
of~$\sim10^5$. 

In this paper, we present coronagraphic performance obtained at the
Palomar telescope using a self-coherent camera. In
section~\ref{sec:scc}, we remind the principle of the self-coherent
camera. We then explain how it was implemented in the stellar double
coronagraph instrument \citep{mawet14b,bottom16} in
section~\ref{sec:sdc}. Finally, after presenting the procedures for
non-common path aberration calibration in section~\ref{sec:ncpa}, we
present the performance of the self-coherent camera on internal source
and on-sky in sections~\ref{sec:inter} and ~\ref{sec:onsky}.

\section{Principle of the Self-coherent camera}
\label{sec:scc}

The performance of coronagraphs is limited by phase and amplitude
aberrations of the wavefront upstream the focal plane mask. In
ground-based telescopes, the adaptive 
optics~(AO) system compensates for most of the atmospheric turbulence
but it cannot provide an aberration-free wavefront to the
coronagraph. Moreover, the~AO estimates aberrations in the wavefront
sensing channel that is different from the science channel. Thus,
non-common path aberrations~(NCPA) are seen by the coronagraph, which
induce stellar speckles that mimic an exoplanet image in the science
image. In space, optical 
aberrations vary because of variations of thermal or gravitational
flexures. Hence, aberrations need to be calibrated regularly during the
observations to have the coronagraph work in optimal conditions. To
avoid~NCPA or varying aberrations, the only efficient techniques --
focal plane wavefront sensors -- estimate the aberrations from the
science image. Doing so means measuring the electric field associated
with the stellar speckle in that plane. To do so, \citet{bottom17}
implemented a phase-shifting interferometer on the~SDC to spatially
modulate the speckle intensity. In this paper, we present the results
obtained when implementing a self-coherent camera~(SCC).

\begin{figure*}[!ht]
  \centering
  \includegraphics[width=.5\textwidth]{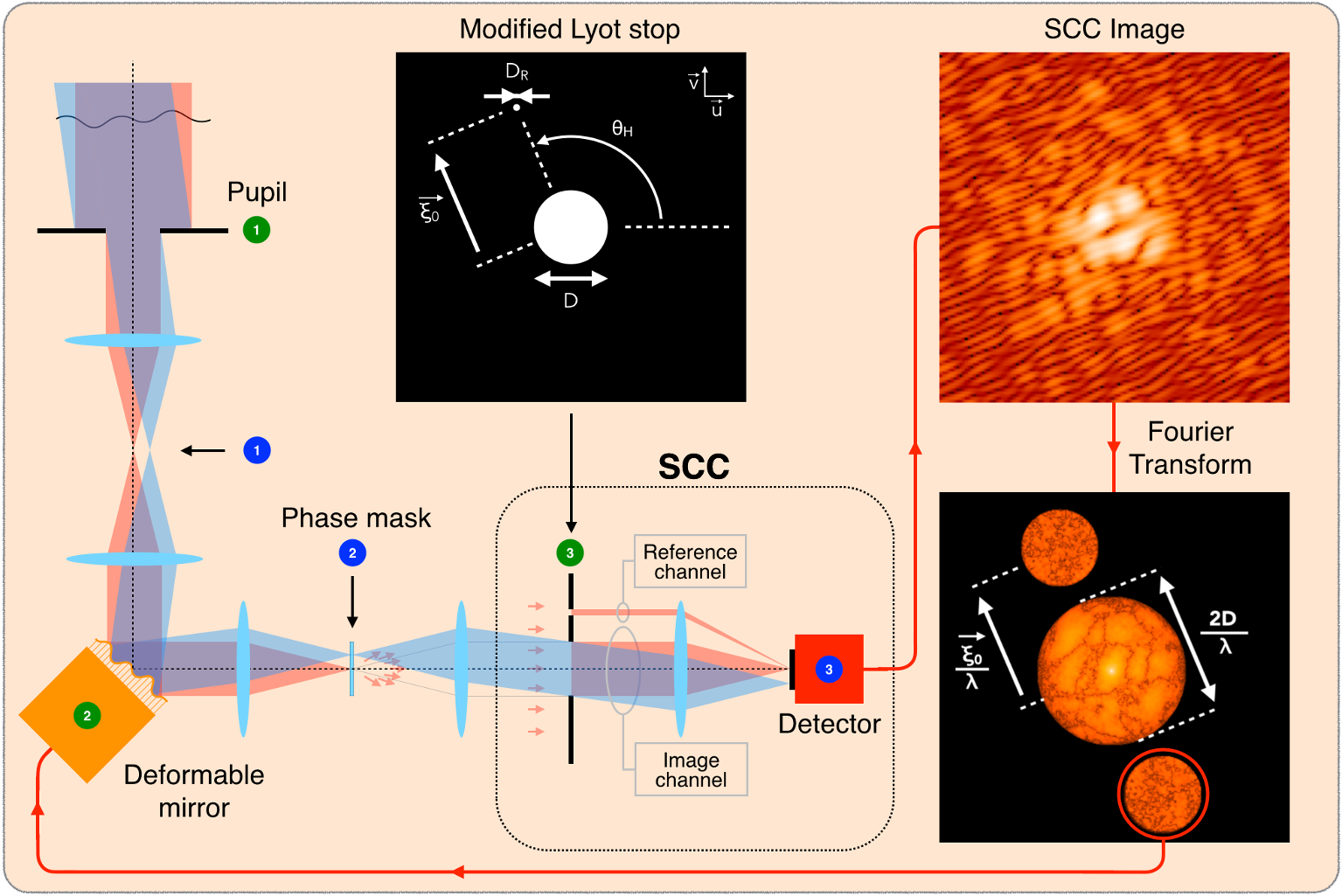}
  \caption{Principle of the self-coherent camera (SCC). The stellar
    beam (red) hits a deformable mirror, goes through a focal plane
    coronagraphic mask and a Lyot stop. Because of wavefront
    aberrations, part of the starlight goes through the image channel inducing
    speckles on the detector. Selecting part of the starlight rejected
    by the focal mask (reference channel), one spatially modulates the
    speckle intensity in the SCC image (top right). It is then possible to
    measure the speckle electric field from the Fourier transform
    of the SCC~image (bottom right) in order to control the deformable
    mirror. The planet light (blue) is not affected by the 
    coronagraph and the planet image is not fringed.}
  \label{fig:fig1}
\end{figure*}

The SCC is a focal plane wavefront sensor that
spatially modulates the intensity of the stellar speckles to retrieve
the associated complex electric field~\citep{baudoz06,
  galicher08, baudoz10b, baudoz12, galicher10}. It was optimized in
laboratory in space-like
conditions~\citep{mazoyer13a,mazoyer14a,galicher14,baudoz18b} and in
ground-based conditions (Singh et al. in Prep). The principle is
recalled in figure~\ref{fig:fig1}.  
The stellar beam (red) hits a deformable mirror. Then, it is focused
onto a coronagraphic focal plane mask that scatters light in the Lyot
stop plane outside the geometrical pupil. A Lyot diaphragm stops the stellar
light before it reaches the detector. If optical aberrations exist,
part of the starlight is scattered inside the Lyot diaphragm and
reaches the detector forming speckles that mimic an exoplanet
image. The~SCC consists on adding a 
small hole in the Lyot stop (top image) to create a 
reference beam that interferes with the image channel and spatially
modulates the stellar speckles in the science image recorded by the
detector (top right). The~SCC is then doing classical off-axis
holography and the lateral peak in the Fourier transform
of that~SCC image (bottom right) provides a direct estimation of the
electric field in the science image. An interaction matrix is then
recorded and it is possible to control a deformable mirror to minimize
for the speckle intensity enhancing the contrast in the science image
\citep{mazoyer13a,mazoyer14a}. If an exoplanet (blue beam) orbits the
targeted star so that its image is not centered onto the focal plane
mask, none of its light goes through the reference channel and its
image is not fringed.

\section{Implementation in the stellar double coronagraph instrument}
\label{sec:sdc}
\begin{figure*}[!ht]
  \centering
  \includegraphics[width=.5\textwidth]{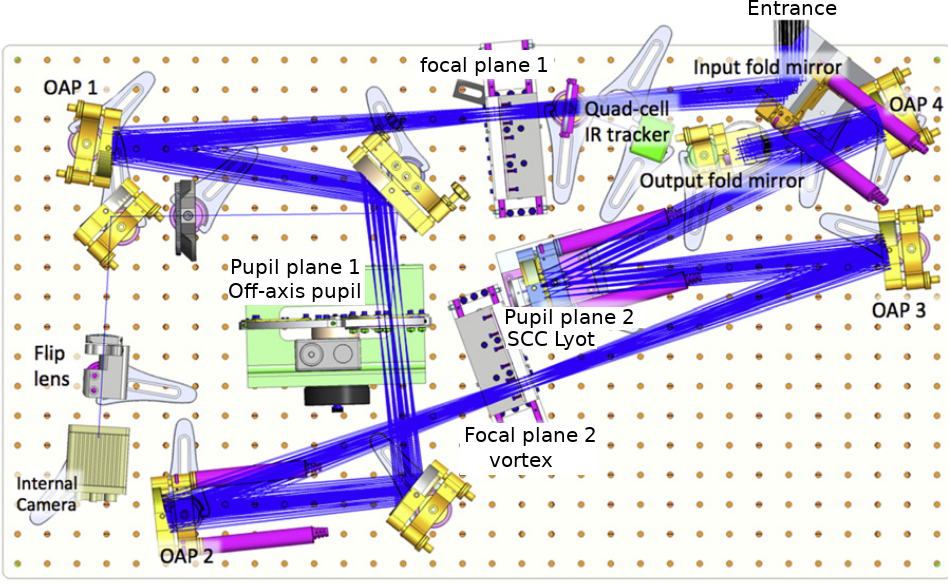}
  \caption{Stellar Double Coronagraph (SDC) instrument layout. In
    Pupil plane~1, an off-axis pupil is set to simulate an off-axis
    $1.5$\,m telescope. A vortex phase mask is set in Focal plane~2 and
  a modified Lyot stop is installed in Pupil plane~2. The beam is then
  sent towards the Pharo detector.} 
  \label{fig:fig2}
\end{figure*}

The stellar double coronagraph (SDC) instrument
\citep[figure\,\ref{fig:fig2}, ][]{mawet14b,bottom16} is
installed at the primary focus of the Hale~200\,inch telescope at
Palomar. It is fed by the~PALM-3000 adaptive optics
system~\citep{bouchez08}. It was designed to cancel the stellar light
using two vortex 
coronagraphs in cascade~\citep{mawet11}. After entering the~SDC bench, the beam goes
through Focal plane~1, Pupil plane~1, Focal plane~2, Pupil plane~2 and
then, it is injected in the PHARO system \citep{hayward01}.

The SCC has already been associated with numerous phase mask
coronagraphs \citep{baudoz18a} reaching very high contrast levels down
to $4.10^{-9}$ in space-like conditions \citep{baudoz18b}. And, as described
in section~\,\ref{sec:scc}, implementing the~SCC is as simple as
adding a small hole in the Lyot stop. This hole diameter
is~$\gamma$ times smaller than the science beam diameter~$D$ and it is
set at more than~$1.5\,D$ from the center of the science beam. Hence, the
optics after the Lyot stop must be twice the science beam so that both
the science and the reference channels can propagate. Such a
constraint is not a problem when designing a new instrument. However,
it forbids the implementation of the~SCC in most of the existing
coronagraphic instruments because the optics after the Lyot stop are
usually a few percents larger than the science beam only.

\begin{figure}[!ht]
  \centering
  \includegraphics[width=.3\textwidth]{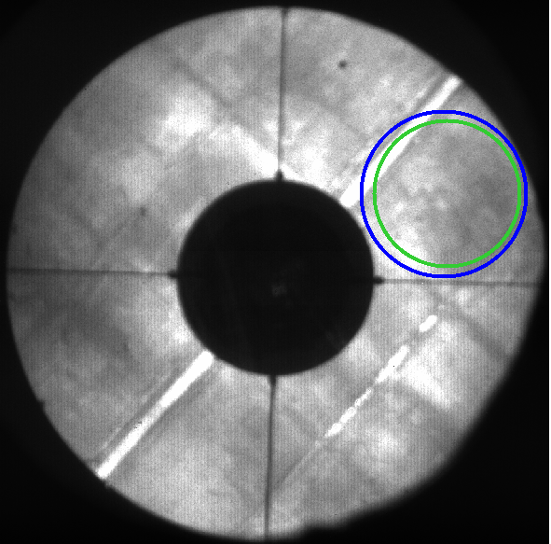}
  \caption{Measured intensity in pupil plane when observing the dome
    of the Palomar observatory through the 5m Hale
    telescope+SDC+Pharo. The blue circle represents the 1.5m off-axis
    pupil that is selected in the first pupil plane in~SDC. The green
    circle shows the position of the Lyot stop. The intensity pattern is
    the dome structure. The spiders are the thin black horizontal and
    vertical lines.} 
  \label{fig:fig3}
\end{figure}

\begin{figure}[!ht]
  \centering
  \includegraphics{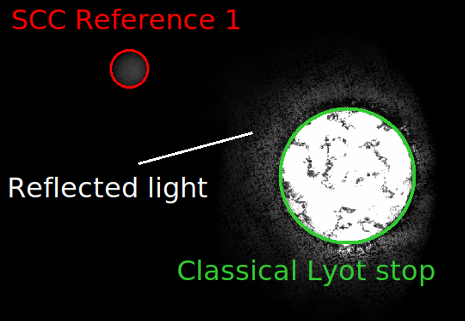}
  \caption{Measured intensity in the Lyot stop plane when the beam of
    the internal source is centered on the vortex coronagraph. The
    classical Lyot stop is encircled in green. The SCC reference hole
    used during the run is encircled in red. The rest of the image
    should be dark as the Lyot mask should stop the star
    light. However, the mask is not perfectly black and reflected light is
    detected close to the Lyot stop.}
  \label{fig:fig4}
\end{figure}

To overcome this limitation and implement the~SCC in the~SDC
instrument, we put no  optics in focal plane~1
(figure\,\ref{fig:fig2}). Doing so, the Hale pupil is
reimaged in pupil plane~1 (figure\,\ref{fig:fig3}).
 There, we add a diaphragm (represented by the blue circle)
 to create a 1.5m~off-axis pupil from the full 5m~obscured
 pupil. Then, we use the vortex phase mask of charge~2 in focal
 plane~2. And in pupil plane~2, we set up a reflective modified Lyot
 stop. The on-axis diaphragm stops the stellar light that is
 scattered by the vortex mask outside the geometrical pupil (classical
 coronagraphic Lyot stop in figure\,\ref{fig:fig4}). The Lyot stop
 diameter is 88\,\% of the off-axis pupil diameter to remove the light
 scattered near the border of the geometrical pupil.
The~SCC reference hole encircled in red in the figure has a
diameter~$\gamma=4$ times smaller than the Lyot stop. As
  explained in~\citet{mazoyer13a}, the light of the reference beam
  mainly spreads in an Airy pattern with a radius
  of~$\sim1.2\,\gamma\,\lambda/D$ in the coronagraphic image. 
  Speckles are thus fringed up to~$\sim1.2\,\gamma\,\lambda/D$ from the
  optical axis. Therefore, the smaller~$\gamma$ the larger the
  field-of-view that can be corrected from speckles. However, the
  larger~$\gamma$ the fainter the intensity of the reference beam in
  the coronagraphic image and the fainter the fringe visibility. A
  trade-off has to be chosen between the fringe visibility and the
  size of the corrected 
  field-of-view. In the case of the~SDC, there was large aberrations
  meaning bright speckles during our run. We had to
  use~$\gamma=4$ so that speckles close to the optical axis were
  correctly fringed. We could then minimize the speckle intensity
  within~$5\,\lambda/D$ from the optical axis. In order to enlarge the
region of correction, we would have reduced the size of the reference
hole (increasing $\gamma$) but we did not due to lack of time.

   In figure\,\ref{fig:fig4}, all but the Lyot stop and the reference
   disks should be dark as the Lyot mask should stop the star
   light. However, the mask is not perfectly black and reflected light is
   detected close to the Lyot stop.

\section{NCPA correction procedure}
\label{sec:ncpa}

\subsection{MGS algorithm limitation}
\label{subsec:mgs}
At the Hale telescope, the current procedure used to minimize~NCPA
before the observations is based on a modified Gertzberg–Saxton~(MGS)
algorithm\,\citep{burruss10}. This technique estimates and minimizes
the phase aberrations recording a set of out-of-focus
non-coronagraphic images. The estimated aberrations include
aberrations upstream and downstream the coronagraphic focal plane
mask, and they are both compensated by a deformable mirror that is
upstream the mask. In the focal plane where the coronagraphic mask is,
the aberrations are thus overcorrected. And, after an~MGS minimization
of~NCPA, part of the stellar light leaks through the coronagraph and
induces stellar speckles in the science image (see
section~\ref{sec:inter}).

\subsection{SCC procedure}
\label{subsec:scc}
To optimize the minimization of the speckle intensity in the science
images, we used the~SCC implemented in the~SDC (section~\ref{sec:sdc})
in closed-loop controlling the~$66\times66$ Xinetics deformable mirror
of the PALM-3000. As the~SCC reference hole that we
used is~$\gamma=4$ times smaller than the Lyot diaphragm, the visibility of
the SCC fringes was detectable up to~$\sim5\,\lambda/D$ from
the star in the science image\,\citep{mazoyer13a}. Hence, we tried to
minimize the speckle intensity up to~$5\,\lambda/D$. To correct at
larger separations, one would need to use a smaller reference hole so
that the reference intensity nulls further away.

One iteration of correction consists on recording
one~SCC coronagraphic science image, estimating the electric field
associated with the stellar speckles from this image
(section~\,\ref{sec:scc}), and sending commands to the deformable
mirror using the control matrix. The control matrix is the
pseudo-inverse of the interaction matrix that we recorded using the
internal source of the~SDC instrument. To record this matrix, we used a
truncated Fourier basis\,\citep{mazoyer14a} composed of all sine and
cosine functions in the pupil plane that induce speckles
below~$5\lambda/D$ in the science image. One row of the interaction
matrix is the estimated electric field in the science image when
applying one function of our basis. To apply the sine/cosine phase
functions, we modified the voltages of the deformable mirror.
  Recording this matrix with the Pharo detector was taking
  about~$20\,$minutes. Once
this calibration was done, we could close the~SCC correction loop to
minimize the~NCPA. Note that we assume small aberrations when
recording the interaction matrix. As a consequence several iterations
of correction are needed to minimize the speckle intensity.

\section{Internal source}
\label{sec:inter}

\subsection{Performance}
\label{subsec:perf}
\begin{figure*}[!ht]
  \centering
  \includegraphics{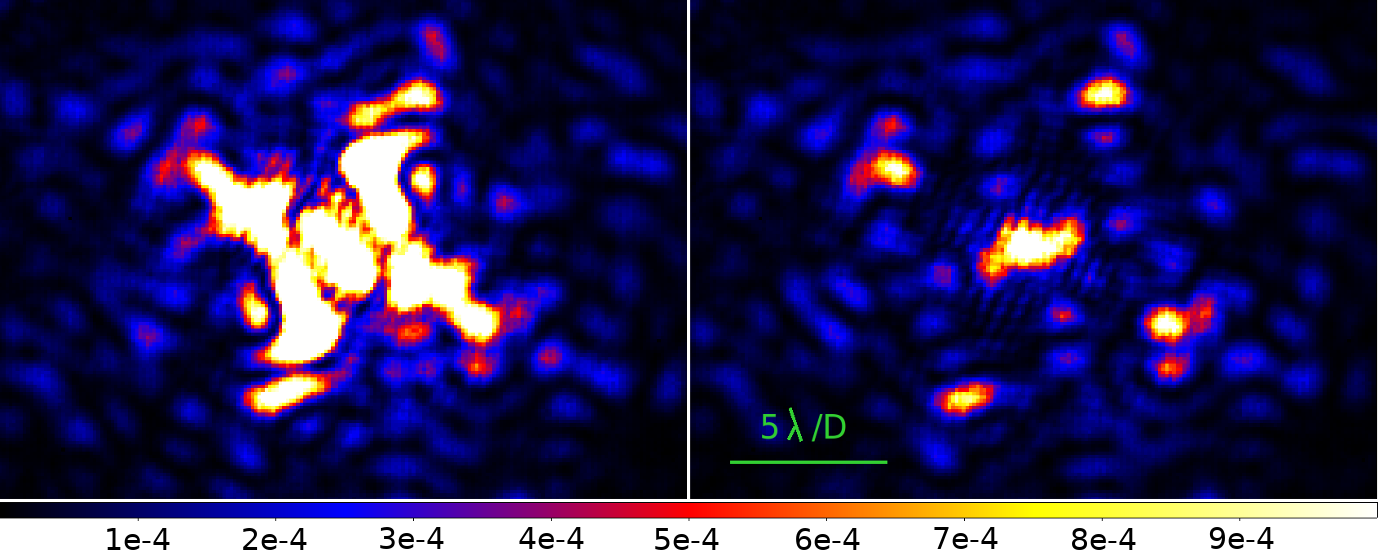}
  \caption{Internal source: science images recorded at~Br-$\gamma$
    after minimization of~NCPA using the~MGS algorithm (left-hand panel) and
    using the~SCC (right-hand panel). Same color scale and same field
    of view for both images. The color bar gives the intensity
    normalized by the non-coronagraphic PSF maximum.} 
  \label{fig:fig5}
\end{figure*}

We were granted two nights at the Hale telescope on the Stellar Double
Coronagraph
(program~3660\footnote{\tiny https://reservations.palomar.caltech.edu/observing\_schedule/abstract/3660}). During daytime on the 25th of July 2018, we
first minimized the~NCPA in 
the~SDC instrument using the~MGS algorithm using the~Br-$\gamma$
filter \citep[$\lambda_0=2.166\,\mu$m
  and~$\Delta\lambda=0.020\,\mu$m,][]{hayward01} and the internal
source. NCPA were reduced but speckles were still present in the~MGS
coronagraphic image (left-hand panel in figure~\ref{fig:fig5}) because the~MGS
solution over-corrects the aberrations upstream the coronagraph focal
plane mask as explained in section~\ref{subsec:mgs}. 

We then used the~SCC to optimize the speckle minimization. We started
from the~MGS image (left-hand panel in figure~\ref{fig:fig5}) in which the~SCC
fringes are detected (from bottom left to top right). We estimated and
corrected the aberrations up to~$5\,\lambda/D$ around the optical
axis. After three iterations, the speckle intensity inside the control
area was efficiently reduced as showed on the right-hand panel in
figure~\ref{fig:fig5}. The four satellite speckles are also
present in the MGS image. They are between~$\sim5\,\lambda/D$
and~$\sim6\,\lambda/D$ and then, outside the corrected area. That is
why they are not corrected in the SCC image.

\begin{figure}[!ht]
  \centering
  \includegraphics{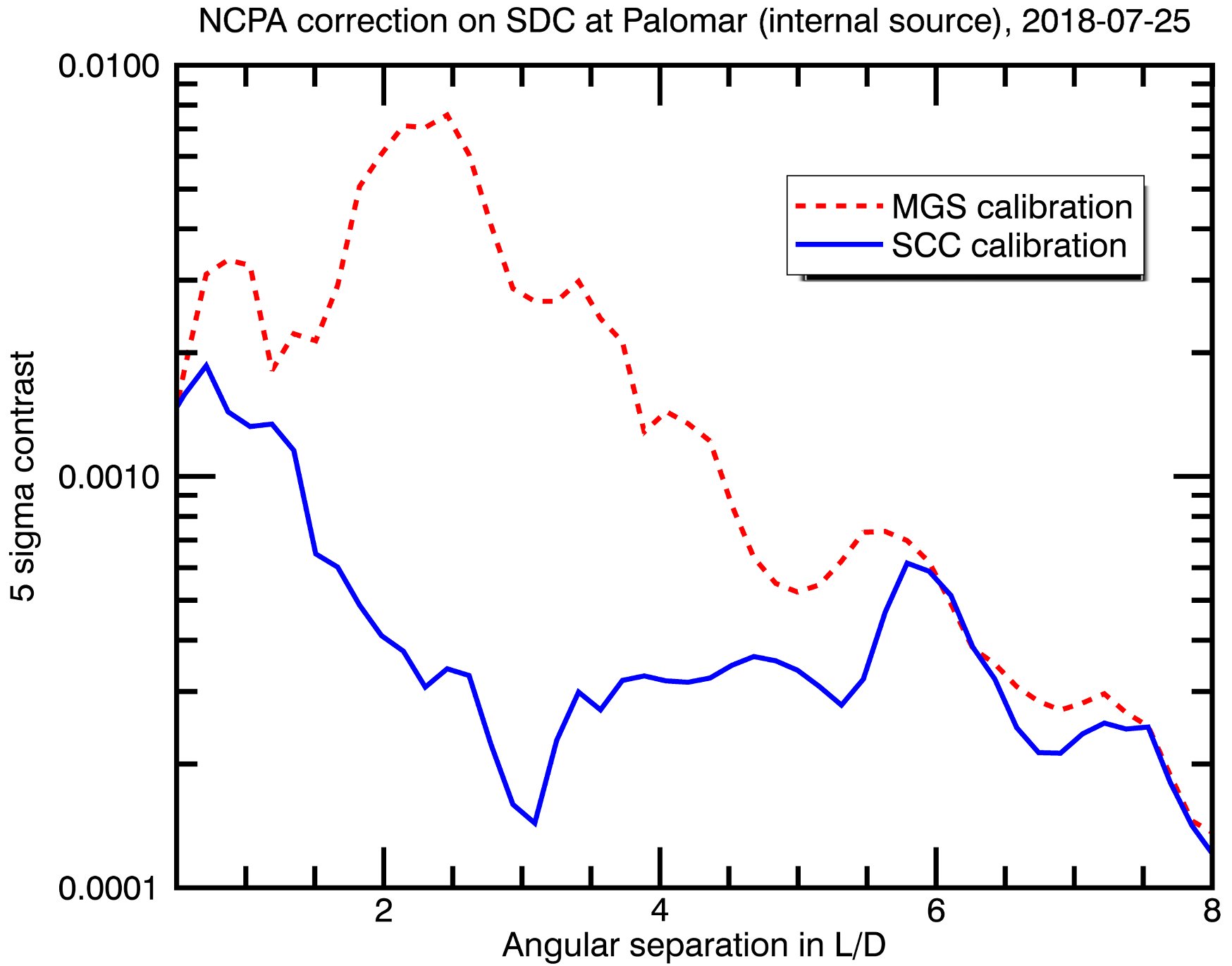}
  \caption{Internal source: $5\,\sigma$ detection limit associated with
    MGS (red dashed line) and~SCC (blue full line) calibrated images of
    figure\,\ref{fig:fig5}.}
  \label{fig:fig6}
\end{figure}

The~$5\,\sigma$ detection limit for the~MGS and the~SCC calibrated
images are plotted in figure~\ref{fig:fig6}.
The detection limit is
the azimuthal standard deviation of the intensity calculated in
annuli of $1\,\lambda/D$ width and centered on the optical axis
(i.e. the star image). The self-coherent camera improves
     the detection limit in the coronagraphic image by a factor
     between~$4$ and~$20$ between~$1.5$ and~$5\,\lambda/D$. Such a
     result demonstrates the efficiency of the~SCC very close to the
     optical axis where many exoplanets are to be discovered and
     where other techniques like~angular and spectral differential
     imaging cannot calibrate the
     speckles~\citep{marois06,racine99}. 

     \subsection{Comparison with other techniques}
     Other techniques
     like reference differential imaging \citep{ruane19} can partially
     calibrate the speckles close to the optical axis. They are
     however limited by speckles with lifetime shorter than few
     minutes (the speckle pattern then changes between the science
     target and the reference star). Other techniques can do focal
     plane wavefront sensing and correction
     \citep{martinache14,bottom16b,cady13,matthews17,bottom17}. Most
     of them use a temporal modulation of the speckle intensity and
     they cannot calibrate the speckles with lifetime shorter than the
     time needed for calibration (usually 4 to 5 images, meaning few
     minutes for a typical star magnitude).

     \subsection{Current limitation at Palomar}
     The contrast level reached in the SDC images (a
     few~$10^{-4}$) is quite moderate when compared to results
     obtained in laboratory \citep[down to
       $4.10^{-9}$,][]{baudoz18b}. Close to the optical axis, the
     performance is set by the SDC vortex phase mask, the reflected
     light by the Lyot stop (see the end of section~\ref{sec:sdc}),
     and the jitter stability during the exposure that is~$\sim1.5\,$s
     at minimum (plus~$\sim6.5\,$s of overhead). We note however that
     the contrast we obtained in section~\ref{subsec:perf}
     ($\sim2.10^{-3}$) is better than the one previously reached using 
     the same 1.5m off-axis configuration \citep[$5.10^{-3}$ to
       $10^{-2}$ in][]{serabyn10}.

     Further away from the optical axis, we believe that the performance can
     be improved in the current SDC instrument reducing the reflected
     light by the Lyot stop and optimizing the SCC speckle calibration
     but we had a limited amount of time at the telescope. However,
     improving the performance in SDC images does not mean reaching
     $4.10^{-9}$ contrast levels. Such a performance is possible using
     a coronagraph that reaches contrast level of a few~$10^{-5}$ on
     the optical axis in a system that remains stable during the
     speckle calibration (gravity and thermal flexures inducing a
     jitter smaller than $\sim\lambda/(10\,D)$). This would be
     possible using a faster detector than Pharo whose highest rate is
     about~$0.1\,$Hz. 
     
\section{Performance on sky}
\label{sec:onsky}
 During our stay at the Palomar Observatory, the quad-cell
  infrared tracker that is used to stabilize the star image on the
  center of the vortex phase mask \citep[i.e. control of the
    jitter,][]{bottom16} was not in service. Therefore, it was not
  possible to close the SCC loop on sky. Moreover, even if the
  quad-cell tracker was used, a faster detector than Pharo would be a
  key point to control the speckles before they evolve because of
  gravity or thermal flexures.

  As we could not
  close the loop on sky, we used another approach. On the 25th of
July 2018, we minimized the speckle intensity on
internal source during daytime using the~MGS algorithm and the~SCC
technique. Between~$3$\,h and~$4$\,h later, we opened the telescope and
pointed Vega with a seeing of~$\sim1.7$\,arcsec. We recorded sequences
of images in~Br\,$\gamma$ with the beam aligned on the vortex
coronagraph.

\begin{figure*}[!ht]
  \centering
  \includegraphics{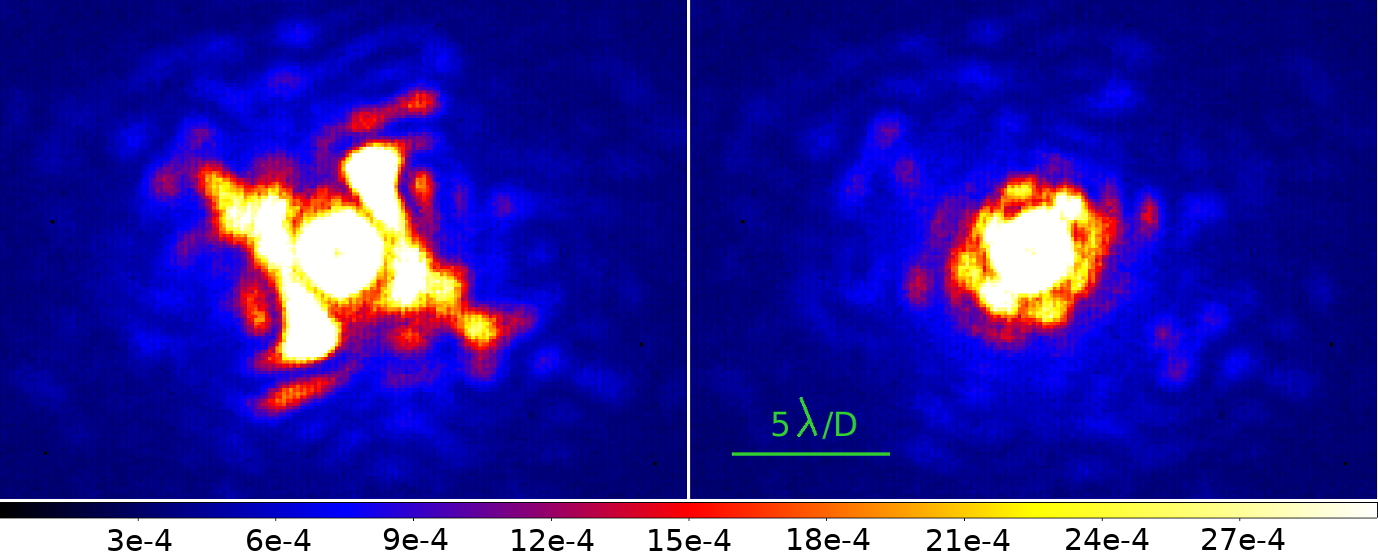}
  \caption{On-sky: science SCC coronagraphic images at~Br-$\gamma$
    using the~MGS solution (left-hand panel) or the~SCC solution
    (right-hand panel). Same
    color scale and same field of view for both images. The color bar
    gives the intensity normalized by the non-coronagraphic PSF
    maximum.}
  \label{fig:fig7}
\end{figure*}

During the first sequence, we applied the~MGS calibration described in
section~\ref{sec:inter}. We recorded~$40$ exposures of
$1.416$\,s. The average of the coronagraphic images is showed on the
left-hand panel in~Fig.~\ref{fig:fig7}. The image is very similar to the one
measured with the internal source except close to the center. This
difference is due to uncorrected jitter when on-sky.

\begin{figure}[!ht]
  \centering
  \includegraphics[width=0.45\textwidth]{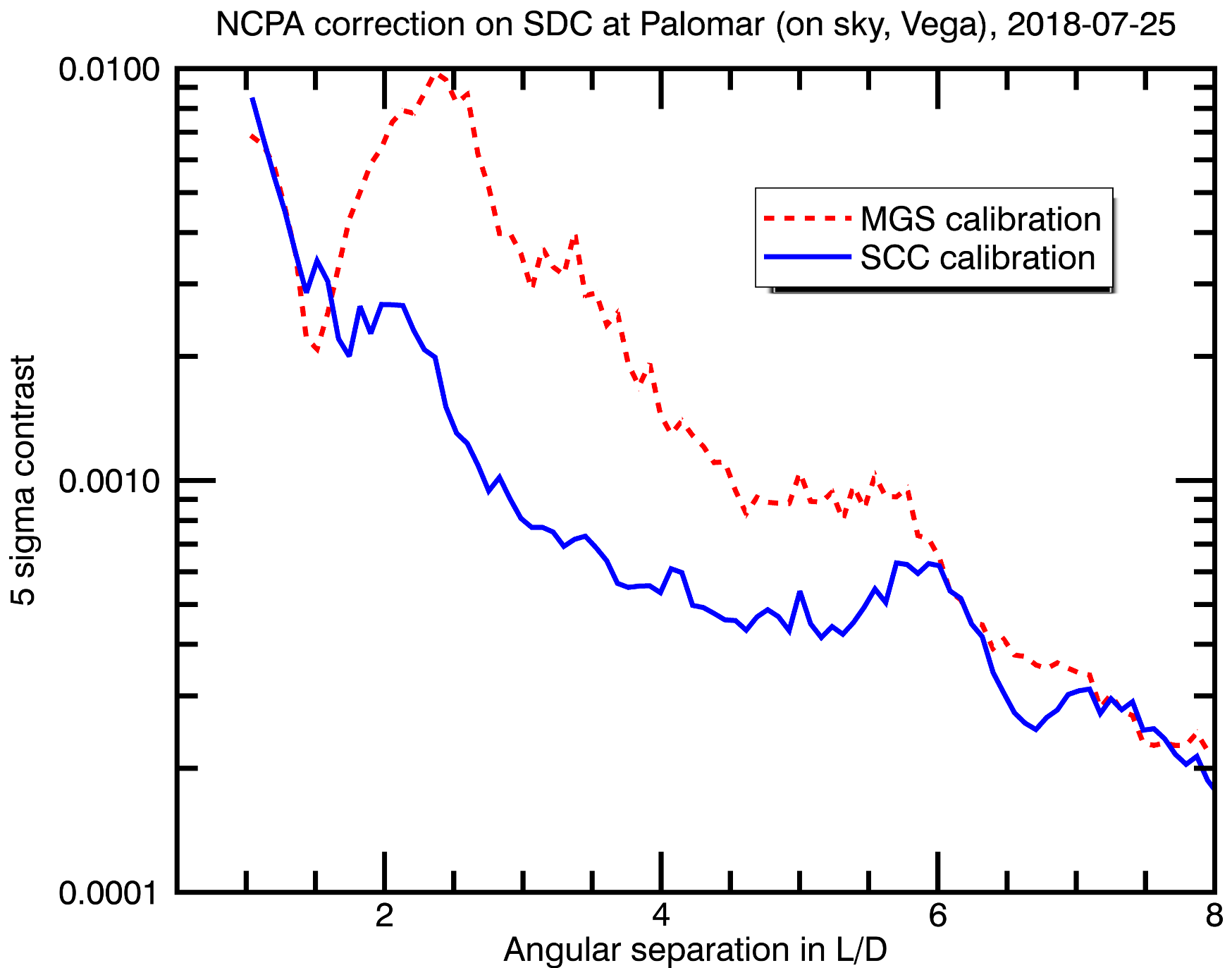}
  \caption{On-sky: $5\,\sigma$ detection limit associated with
    MGS (red dashed line) and~SCC (blue full line) calibrated images of
    figure\,\ref{fig:fig5}.}
  \label{fig:fig8}
\end{figure}

During the second sequence, we applied the~SCC solution obtained with
the internal source and we recorded~$20$ exposures of~$1.416$\,s. The
averaged image is showed on the right-hand panel in~Fig.~\ref{fig:fig7}. As with
the internal source, the quality of the image is clearly improved with
respect to the~MGS solution. Speckles are suppressed from the
image. This is confirmed when plotting the~$5\,\sigma$ detection limit
(Fig.~\ref{fig:fig8}).
Between $2$ and~$4\,\lambda/D$, the detection limit is~$\sim5$ times
better after~SCC calibration than after~MGS calibration. The
difference with the image obtained on internal source
(section~\ref{sec:inter}) are the starlight leakage close the
star center due to uncorrected jitter, a smooth halo created by
averaging turbulence speckles, and residual speckles above this halo
up to~$\sim5\,\lambda/D$. The latter can be
induced by uncorrected static aberrations due to telescope
aberrations that were not calibrated using the internal source.

\section{Conclusion}
To improve exoplanet imaging instruments, a crucial point is to
actively compensate for the non-common path aberrations (NCPA) between
the classical adaptive system channel and the science coronagraphic
image. The use of a focal plane wavefront sensor is optimal to
estimate the electric field of the stellar speckles from the
coronagraphic image.  Such a sensor is the self-coherent camera
(SCC) that spatially modulates the speckle intensity. To implement
the~SCC on the existing stellar double coronagraph at the~$200''$
Palomar telescope, we added a small off-axis hole in the Lyot stop of
the vortex coronagraph.

Using the~SCC, we improve the detection limit between~$1.5$
and~$5\,\lambda/D$ by a factor of~$4$ to~$20$ in the laboratory using
the internal source. We then tested on-sky
the quality of the internal calibration. We observed Vega and showed
that the~SCC calibration was~$5$ times better between~$2$
and~$4\,\lambda/D$ when compared to the Palomar standard calibration
used for~NCPA minimization. The loss of performance of the~SCC
calibration between on-sky and on internal source may come from
residual unaveraged aberrations in long but not infinite
exposures. Further telescope time is needed to investigate this
issue.

To conclude, we demonstrated the capacity of the self-coherent camera
to calibrate NCPA in an existing instrument. This first
  demonstration was made using a narrow band filter but implementation
  of the SCC in broadband is possible using several SCC reference
  holes as explained in \citet{delormejr16}. The results related in
this paper also highlight two interests of the~SCC. First, the~SCC
can minimize the speckle intensity in the field of view especially the ones
that are very close to the star where many exoplanets are to be
discovered. Then, even if the SCC requires a~3-4\,pixel sampling on
the science detector instead of~$2$ for other focal plane wavefront
sensors, the~SCC has a~$100\%$ efficiency with science time
as each image can be used for both science and~NCPA
minimization.


\begin{acknowledgements}
The authors thank the {\it R\'egion \^Ile-de-France} and the Science
Council of the Paris Observatory that supported this work.
\end{acknowledgements}

\bibliographystyle{aa}   
\bibliography{aa}

\end{document}